\documentclass[
journal=ancac3, 
manuscript=article]{achemso}
\usepackage{soul}
\usepackage[version=3]{mhchem} 

\author{Everett D. Grimley}
 \email{edgrimle@ncsu.edu}
\affiliation{Department of Materials Science and Engineering, North Carolina State University, Raleigh, NC 27695-7907, USA}
\author{Tony Schenk}
\affiliation{NaMLab gGmbH, Noethnitzer Str. 64, 01187 Dresden, Germany}
\author{Thomas Mikolajick}
\affiliation{NaMLab gGmbH, Noethnitzer Str. 64, 01187 Dresden, Germany}
\altaffiliation{Institute of Semiconductors and Microsystems, TU Dresden D-01062 Dresden, Germany}
\author{Uwe Schroeder}
\affiliation{NaMLab gGmbH, Noethnitzer Str. 64, 01187 Dresden, Germany}
\author{James M. LeBeau}
 \email{jmlebeau@ncsu.edu}
\affiliation{Department of Materials Science and Engineering, North Carolina State University, Raleigh, NC 27695-7907, USA}

\title{Atomic Structure of Domain and Interphase Boundaries in Ferroelectric HfO$_2$}

\begin{document}

\begin{abstract}


\noindent Though the electrical responses of the various polymorphs found in ferroelectric polycrystalline thin film HfO$_2$ are now well characterized, little is currently understood of this novel material's grain sub-structure. In particular, the formation of domain and phase boundaries requires investigation to better understand phase stabilization, switching, and interconversion. Here, we apply scanning transmission electron microscopy to investigate the atomic structure of boundaries in these materials. In particular, we find orthorhombic/orthorhombic domain walls and coherent orthorhombic/monoclinic interphase boundaries formed throughout individual grains. The results inform how interphase boundaries can impose strain conditions that may be key to phase stabilization. Moreover, the atomic structure near interphase boundary walls suggests potential for their mobility under bias, which has been speculated to occur in perovskite morphotropic phase boundary systems by mechanisms similar to domain boundary motion.

\end{abstract}

\section{Introduction}

Following the first report of its ferroelectric behavior,research on hafnia (HfO$_2$) systems has continued with renewed interest \cite{boscke_ferroelectricity_2011}. This is spurred on by its robust, lead-free ferroelectric properties that are maintained even in films thinner than 10 nm. Because of its silicon compatibility and wide processing space, the material shows promise for use in future memories \cite{muller_ferroelectricity_2012,pesic_nonvolatile_2016}, energy efficient logic transistors \cite{mulaosmanovic_evidence_2015,hoffmann_direct_2016}, and devices that exploit a tunable dielectric \cite{dragoman_extraordinary_2017} or pyroelectric \cite{hoffmann_ferroelectric_2015,park_giant_2016,smith_pyroelectric_2017}.

As understanding of ferroelectricity in hafnia develops, grain sub-structure is proving increasingly important for controlling film properties. While bulk hafnia is known to adopt the \textit{P$2_1$/c} monoclinic phase (M) at room temperature and pressure \cite{adams_x-ray_1991}, ``metastable'' high symmetry fluorite-like phases including \textit{P$4_2$/nmc} tetragonal (T) and orthorhombic (O) phases can coexist in the ferroelectric thin films \cite{boscke_ferroelectricity_2011,muller_ferroelectricity_2012,sang_structural_2015,richter_fragile_2017,smith_pyroelectric_2017,park_comprehensive_2017}. An orthorhombic phase \textit{Pca2$_1$} that lacks an inversion center is thought to be responsible for the ferroelectric behavior of these thin films, and has been observed with scanning transmission electron microscopy (STEM) \cite{sang_structural_2015,shimizu_growth_2015}. Electron microscopy has also revealed interfacial hafnia regions exhibiting tetragonal-like symmetry at electrode/bulk grain interfaces in moderately doped films \cite{pesic_physical_2016,grimley_structural_2016,richter_fragile_2017}, and its presence dominates at high dopant concentrations \cite{park_comprehensive_2017,richter_fragile_2017}. Critically, the net electrical behavior is strongly governed by the fractions of each phase in a given device \cite{park_effect_2016,richter_fragile_2017,park_comprehensive_2017}.

First-principles calculations suggest that various forces contribute to stabilizing the different distorted fluorite phases of HfO$_2$, enabling ferroelectric switching, and/or possibly allowing phase transformation. These include electric fields \cite{materlik_origin_2015,batra_factors_2017}, surface energies \cite{materlik_origin_2015,batra_stabilization_2016}, strain from different origins \cite{reyes-lillo_antiferroelectricity_2014,materlik_origin_2015,batra_factors_2017,barabash_ferroelectric_2017}, and alloying \cite{materlik_origin_2015}. Experiment and theory point to an orthorhombic switching pathway through the tetragonal phase  \cite{muller_ferroelectricity_2012,huan_pathways_2014,reyes-lillo_antiferroelectricity_2014,barabash_ferroelectric_2017}, and in certain instances the tetragonal-to-orthorhombic transition might be transient during the application of an electric field \cite{pesic_nonvolatile_2016}. 

Recently, studies have also highlighted the structural similarities between the orthorhombic and monoclinic phases \cite{clima_identification_2014,barabash_ferroelectric_2017}. Barabash et al.~report that differences in oxygen ordering in a ``parent'' orthorhombic phase (centrosymmetric \textit{Pbcm}) can lead to stabilization of either the monoclinic or the polar orthorhombic phase. Furthermore, they speculate that a region of coherently strained HfO$_2$ lacking the monoclinic distortion might readily convert between the monoclinic and polar orthorhombic phase via a low transformation barrier \cite{barabash_ferroelectric_2017}. Experimental evidence also suggests that some amount of phase transformation may occur during the ``wake-up'' effect \cite{martin_ferroelectricity_2014,park_study_2015,kim_study_2016,pesic_physical_2016,grimley_structural_2016}. The complexities of characterizing  polycrystalline and polyphasic hafnia thin films have, however, limited current information of phase distribution, coexistence, and domain structuring in this new ferroelectric system.


Internal boundaries are also crucial to consider as they can impact a ferroelectric material's mechanical and electrical response. This has been seen, for example, near morphotropic phase boundaries (MPBs) in the phase diagrams of certain materials.  Pb(Zr,Ti)O$_3$ exhibits coexistence of polar rhombohedral and polar tetragonal phases, which exist in fractions and over length scales that depend largely on the composition \cite{glazer_influence_2004}. Domain wall energy is an important parameter for determining the length scale of ordering and the domain sizes, and thus has important implications for mechanical and electrical behavior \cite{rossetti_ferroelectric_2008}. Because these systems contain multiple phases, ``interphase boundaries'' can form as walls between different phases. Furthermore, mobile interphase boundaries are speculated to move during cycling in small reversible and irreversible jumps like domain walls \cite{damjanovic_chapter_2006,ma_creation_2012,jones_domain_2012}. The presence of interphase boundaries in ferroelectric hafnia would thus be expected to influence its phase stability through internal strains at immobile boundaries, and its electrical properties in the case of those that are mobile which would contribute to changes in phase fractions. 

In this article, interphase boundaries and single phase domains in Gd doped HfO$_2$ metal-ferroelectric-metal capacitors are studied using aberration corrected scanning transmission electron microscopy (STEM). Monoclinic, orthorhombic, and tetragonal regions are found to coexist within single grains, and the presence of domain walls in the orthorhombic phase are correlated to field-cycling history. Monoclinic/orthorhombic interphase boundaries are revealed and analyzed in the context of the structural parameters that govern their formation.  Moreover, our results highlight the similarities between the orthorhombic and monoclinic phases. These similarities lead to challenges in distinguishing a ``defect'' in one phase from the ``normal'' structure of the other phase. These combined results suggest that the environments near interphase boundaries lead to the formation of new orthorhombic regions.  Contingent on the stability/mobility of these boundaries, such boundaries are proposed to facilitate some degree of phase conversion under electrical bias.

\section{Results and Discussion}

\begin{figure}[htbp]
	\includegraphics[width=3.25in]{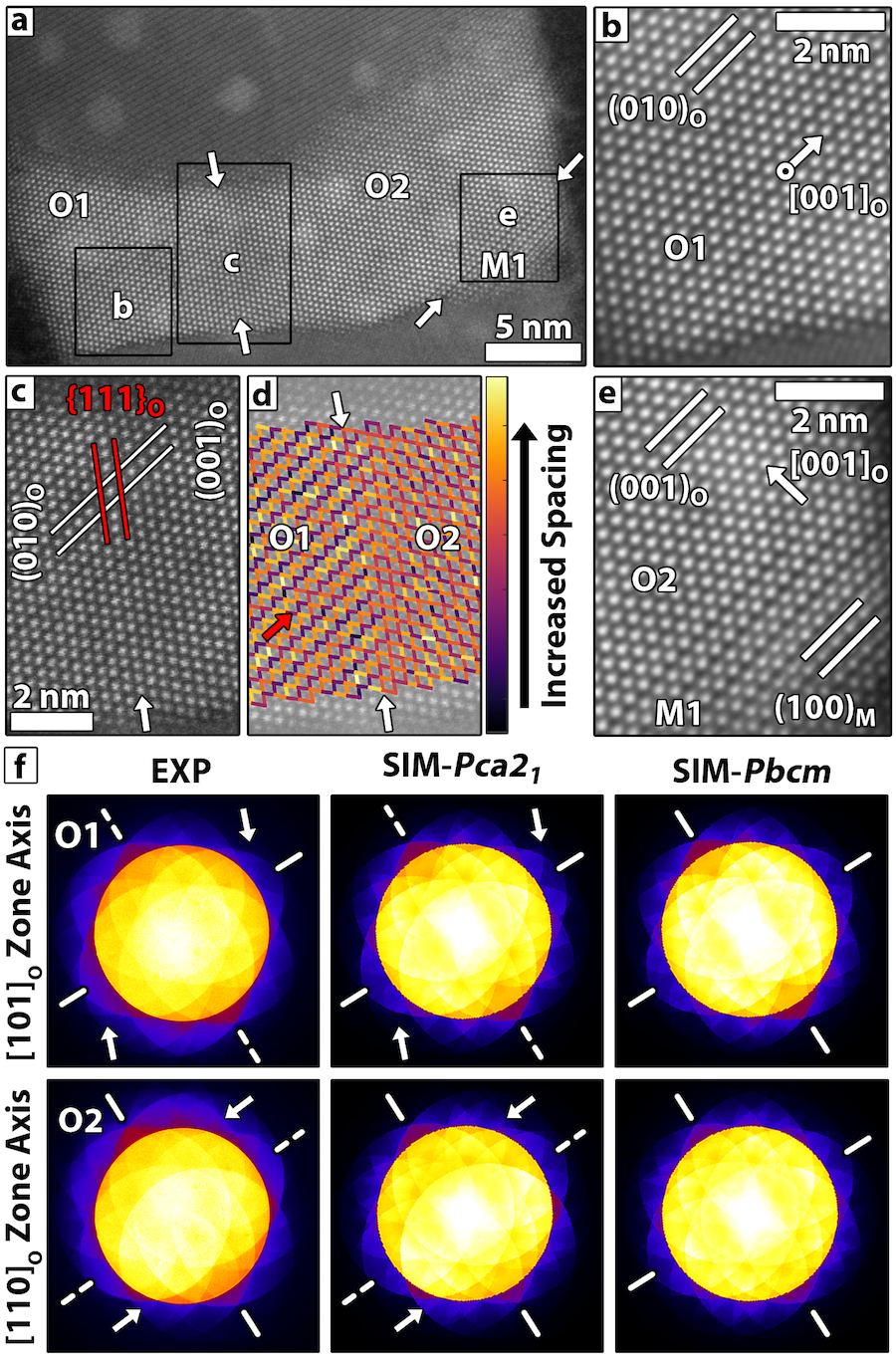}
	\caption{HAADF STEM images of: (a) Gd doped HfO$_2$ grain with O and M regions separated by  boundaries indicated by white arrows, (c-d) O1/O2 boundary in (a) and lattice distance map of the  with the red arrow highlighting a change in projected symmetry, and (b,e) magnified regions from (a) where planes are indicated with lines and the polar direction by  arrows. (f) Experiment and simulated PACBED patterns corresponding to O1 and O2 regions. The presence and lack of a mirror plane are solid and dashed lines, respectively. Arrows highlight symmetry breaking in the pattern. Brightness and contrast are adjusted to emphasize PACBED pattern asymmetry.}
	\label{fig:figure1}
\end{figure}

HfO$_2$ grains typically span the thickness of the film between the TiN electrodes, as shown by the bright grain spanning the distance between the two dark electrodes in Figure \ref{fig:figure1}a. Using high-angle annular dark-field (HAADF) STEM, the identity and orientation of phases in HfO$_2$ films are readily determined using the atomically resolved positions of the projected Hf atom sub-lattice \cite{sang_structural_2015,grimley_structural_2016}. This analysis reveals that certain grains exhibit a complex domain structure. For example, a single grain is divided into two orthorhombic (O1, O2) and one monoclinic (M1) region in Figure \ref{fig:figure1}a. At the O1/O2 boundary in Figure \ref{fig:figure1}c,  $\left(010\right)_o$ in O1 are parallel to  $\left(001\right)_o$ in O2, where  $\{111\}_o$ are continuous across the domain wall. The boundary between the two regions is sharp and possesses an abrupt change in projected symmetry at the domain wall. This symmetry change is made more visible by inspection of the atom column near neighbor distances, which are mapped in Figure \ref{fig:figure1}d. An interphase boundary is also observed within the same grain (see Figure \ref{fig:figure1}e), where the crystal structure abruptly transitions from O2 to M1 with $(001)_o\|(100)_m$. 

Connecting structure to polarization is essential for understanding the ferroelectric behavior of HfO$_2$ thin films. Polarization across the O1/O2 domain wall can be assessed by position-averaged convergent beam electron diffraction (PACBED), where missing mirror symmetry in the pattern corresponds to a lack of inversion symmetry in the material \cite{lebeau_position_2010, sang_structural_2015}. This occurs for the \textit{Pca2$_1$} orthorhombic phase along the $\left[001\right]_o$, and is indicated by arrows in Figure \ref{fig:figure1}b,e (note that the $\left[001\right]_o$ in Figure \ref{fig:figure1}b possesses a component out of the image plane).  Figure \ref{fig:figure1}f shows PACBED patterns acquired from regions O1 and O2. Each pattern lacks a mirror plane across the dashed axis bisecting the pattern, which is consistent with the \textit{Pca2$_1$} polar phase  simulations. In contrast, simulated patterns from the centrosymmetric \textit{Pbcm} phase retain mirror symmetry along both axes. These results show that the polar direction is rotated by $\sim90^\circ$, hence forming a 90$^\circ$ domain wall. 

\begin{figure}[H]
	\includegraphics[width=6.5in]{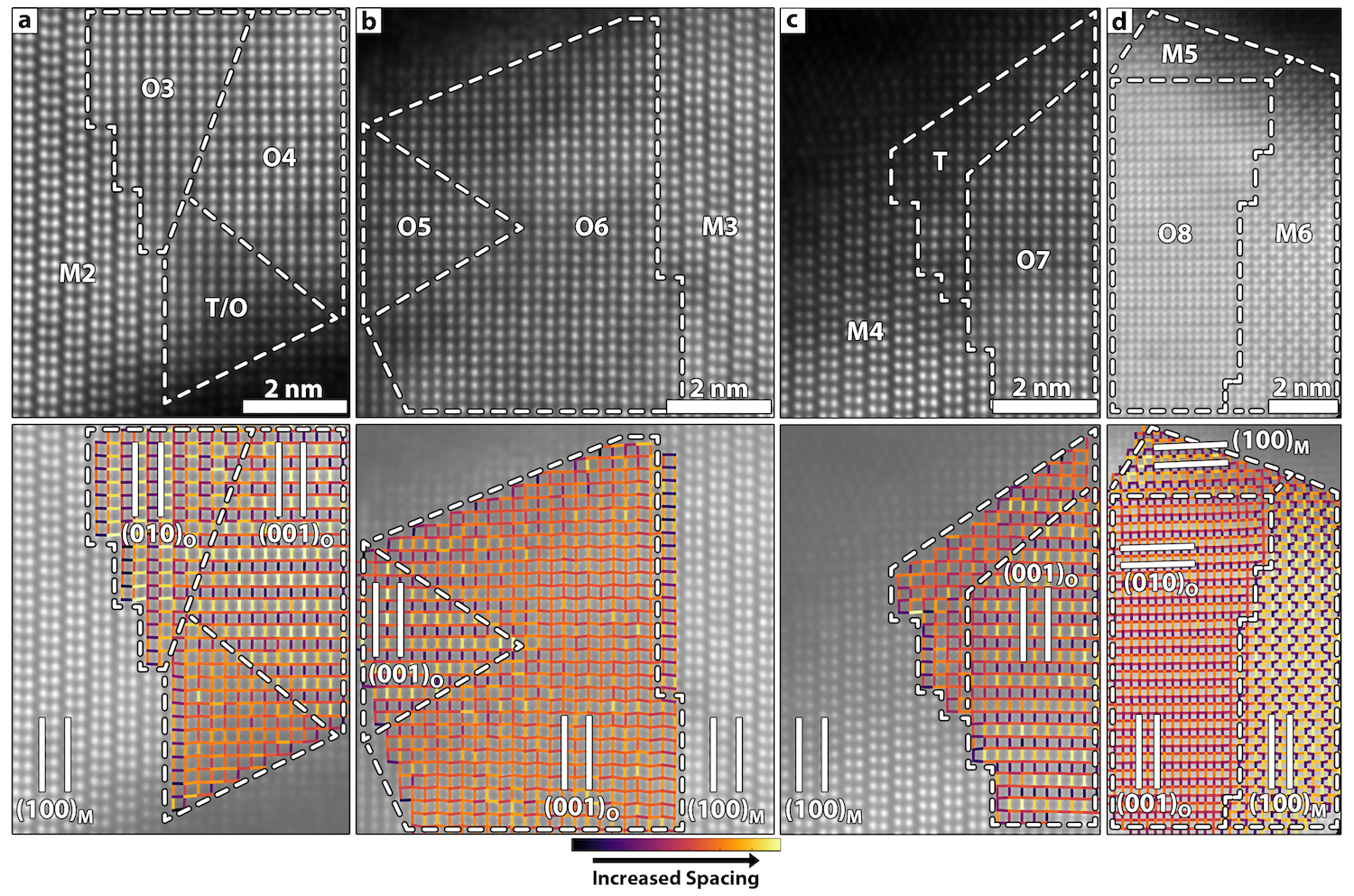}
	\caption{(a)-(d) HAADF STEM images of various regions containing O/O domains and/or O/M interphase boundaries. Dashed lines indicate domain/interphase boundaries. White lines denote indicated Hf planes. Short, colored lines map distances between neighboring Hf sub-lattice atom columns to help guide the eye. Figure \ref{fig:figure2} can be found without overlays in Supporting Information Figure S2.}
	\label{fig:figure2}
\end{figure}


More broadly, a wide range of interphase and domain boundaries are observed throughout the sample as highlighted by Figure \ref{fig:figure2}a-d. The images reveal several O/O domains and O/M interphase boundary structures in a variety of shapes and sizes. Changes in atom column spacing and symmetry across the boundaries are highlighted by near neighbor distance maps in the bottom panels of Figure \ref{fig:figure2}a-d. While some boundaries are angular and difficult to precisely locate, others are flat with fractional unit-cell steps.  Figure \ref{fig:figure2}a shows a typical 90$^{\circ}$ domain between orthorhombic regions O3 and O4. As in the case of the O1/O2 boundary in Figure \ref{fig:figure1}c, $(010)_o$ becomes  $(001)_o$ across the O3/O4 domain wall. The transition in crystal symmetry is abrupt, and like the O1/O2 boundary, occurs over a higher order crystal plane. Not all boundary transitions are sharp in the vicinity of 90$^{\circ}$ domain walls. For example, Figure \ref{fig:figure2}b shows a 90$^\circ$ domain wall formed at the interface between regions O5 and O6. Unlike the O/O domains in Figures \ref{fig:figure1}c and \ref{fig:figure2}a, $(001)_o$ of this domain remain parallel across the boundary and instead rotate 90$^\circ{}$ in-plane. 



Based on the domain walls presented in Figures \ref{fig:figure2}a-b, which are viewed down low order zone axes, some aspects of the structure of orthorhombic domain boundaries can be linked to  misfit. As reported in Ref.~\cite{sang_structural_2015}, the lattice parameters of the orthorhombic structure are \textit{a$_o$,b$_o$,c$_o$} = 5.24, 5.06, 5.07 \AA{}, which results in a $\sim$0.2 \% misfit across the O3/O4 domain wall and a rather abrupt change in rotation across the boundary. In contrast, the O5/O6 boundary exhibits high misfit of $\sim$3.5\%, and possesses a much more diffuse transition in structure across the boundary. This likely arises due to the larger misfit and/or  grain overlap. The three examples of orthorhombic domains in Figures \ref{fig:figure1}c and \ref{fig:figure2}a-b exemplify how local environments allow domains to form in a variety of orientations, sizes/shapes, and domain wall configurations. 

The monoclinic phase is found to form twin boundaries in some grains. For example, a $(110)_m$ twin in Figure \ref{fig:figure2}d is identified between the monoclinic regions M5 and M6. Twining also occurs on $(001)_m$ (Supporting Information, Figure S1a) and $(100)_m$  (Supporting Information, Figure S1b). These twin planes are in good agreement with various reports of twinning configurations identified in toughened zirconia ceramics \cite{bailey_monoclinic-tetragonal_1964}, HfO$_2$ thin films grown directly on Si \cite{maclaren_texture_2009}, and in Hf-rich Hf$_x$Zr$_{1-x}$O$_2$ nanocrystals \cite{tang_martensitic_2005}. Twinning is associated with the tetragonal to monoclinic martensitic phase transformation \cite{bailey_monoclinic-tetragonal_1964,tang_martensitic_2005,maclaren_texture_2009,hudak_real-time_2017}. Such a phase transformation requires a shape change to the distorted monoclinic cell, and twinning is a mechanism whereby shape change/shear strain can be minimized for the transformation of a confined grain \cite{bailey_monoclinic-tetragonal_1964}.


Figure \ref{fig:figure2}a-d also shows that many HfO$_2$ regions contain interphase boundaries. In Figure \ref{fig:figure2}a, an interphase boundary between M2/O3 regions forms with an interface with $(100)_m\|(010)_o$. The interphase boundary wall is discontinuous, with steps forming every few nanometers. Strain near the wall results in visible distortion of the spacing and angle between atom columns in the vicinity of the boundary, as seen in the Figure \ref{fig:figure2}a distance map. Similarly, the structures become blurred adjacent to the boundary wall, which can indicate phase overlap or non-uniform lattice distortion near the boundary. Furthermore, the M4/O7 boundary in Figure \ref{fig:figure2}c forms an interface with $(100)_m\|(001)_o$, and where lattice distortion visible in the vicinity of the interface. Visually, the M4/O7 interphase boundary resembles the M2/O3 boundary, but the orthorhombic region is rotated 90$^{\circ}$ such that $(001)_o$ forms the boundary rather than $(010)_o$. 

In addition to the M/O boundaries in Figure \ref{fig:figure2}a,c mixed tetragonal/orthorhombic symmetry is observed near the TiN electrodes. Interfacial hafnia layers near TiN electrodes are previously reported to relax towards tetragonal symmetry in some instances \cite{grimley_structural_2016}. We propose that these these current findings indicate the local environment near an interphase boundary can promote the stabilization of the tetragonal phase deeper into the grain bulk to $\sim$2-4 nm. Transition regions with mixed/strained symmetry like the tetragonal interface layers reported earlier can be important for phase stabilization \cite{kunneth_modeling_2017}.


The interphase boundaries in Figures \ref{fig:figure2}b,d have reduced step density compared to those in Figures \ref{fig:figure2}a,c. For example at the O6/M3 interface in Figure \ref{fig:figure2}b, an abrupt interphase boundary is formed with $(001)_o\|(100)_m$. The transition in crystal symmetry from orthorhombic to monoclinic is sharp in this case, having less distortion. Further, the boundary between the O8/M5 regions shows no clear steps between the $(010)_o\|(100)_m$ planes that form the wall in Figure \ref{fig:figure2}d. A second boundary in this region also forms between O8/M6 regions with $(001)_o\|(100)_m$, with varied step density. Furthermore, the O6/M3 and O8/M6 interphase boundaries are equivalent, though they are viewed along different crystal projections.


Based on these observations interphase boundaries are expected to traverse  complicated, three dimensional paths through the grain. The nature of the final domain wall thus depends on the size and orientation of the regions that form during annealing. Furthermore, some of the distortion visible near the domain walls is likely the result of viewing a projection of the three-dimensional domain wall structure.  


The library of observed interphase boundaries gives insight into how crystal chemistry may influence their formation.  The examples of interphase boundaries in Figures \ref{fig:figure1} and \ref{fig:figure2} suggests that the monoclinic and orthorhombic phases tend to form coherent boundaries across low order planes in polycrystalline hafnia thin films. This is consistent with phase boundaries seen in strained epitaxial (Hf,Zr)O$_2$ thin films \cite{kiguchi_solid_2016}. Additionally, the boundary step structure suggests a role of misfit in determining their periodicity. The misfit here is defined as the difference in lattice parameters divided by their average. For example, the greatest possible uniaxial misfit  occurs at O/M boundaries where the c$_m$ axis of the monoclinic phase (c$_m$= 5.30 \AA) aligns to either the orthorhombic b$_o$ axis ($\sim$4.6~\% misfit) or c$_o$ axis ($\sim$4.4~\% misfit), where c$_o$ = 5.07 \AA{} and b$_o$ = 5.06 \AA{}. These boundaries with maximum misfit still form and readily step as seen in Figure \ref{fig:figure2}a,c. Comparatively, when O/M boundaries form such that the a$_o$ and c$_m$ axes are parallel, misfit is significantly reduced to  $\sim$2.0~\% when b$_m$/c$_o$ and $\sim$2.2~\% when b$_m$/b$_o$, where b$_m$ = 5.17 \AA{}. These lower misfit boundaries subsequently contain fewer steps as seen in Figure \ref{fig:figure2}b,d.


\begin{figure}[H]
	\includegraphics[width=3.25in]{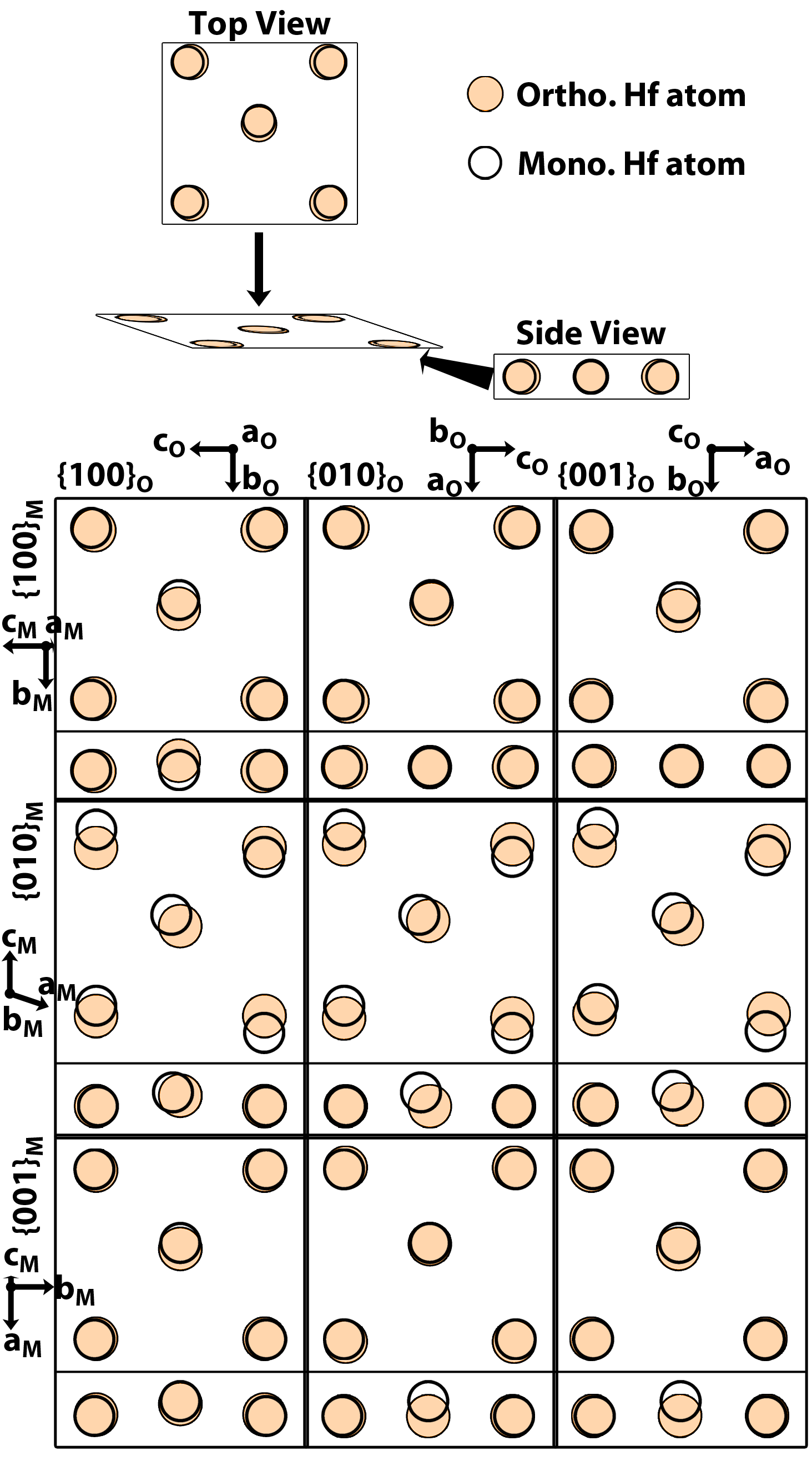}
	\caption{Schematic of a single Hf planes in monoclinic (open circles)  and orthorhombic (filled circles) cells viewed from the top and side.  The O sub-lattice is omitted because it is not observed in the HAADF STEM images. The structures are derived from unit-cell parameters from  Ref.~\cite{adams_x-ray_1991} for the monoclinic cell and Refs.~\cite{kisi_crystal_1989, sang_structural_2015} for the orthorhombic cell. }
	\label{fig:figure3}
\end{figure}


Figure \ref{fig:figure3} shows structural schematics of an interfacial plane, focusing on the Hf positions.  Only a single variant of the boundary is shown here, i.e. rotating the orthorhombic lattice in-plane by 90$^{\circ}$ will change the interface orientation, strain, and alignment of the atoms. Because the $(010)_m$ is non-orthogonal ($\beta$ = 99.18$^{\circ}$), there is relatively poor registration to all orthorhombic planes. In contrast, the set of boundaries formed with  $(100)_m$ and $(001)_m$ provide reasonable registry with the low order orthorhombic planes.


Both orthorhombic and monoclinic cells possess reduced symmetry involving lateral and out-of-plane shifts in atom positions. When viewed from the  side, Figure \ref{fig:figure3}, the Hf sub-lattice remains co-planar for the $(100)_m$, $(001)_o$, and $(010)_o$, while it is rumpled out-of plane for $(010)_m$, $(001)_m$, and $(100)_o$. Based on these observations, interphase boundaries tend to form in orientations that maintain a co-planar Hf sub-lattice across the boundary, i.e.~without out-of-plane rumpling. Furthermore, the local Hf-O bonding configuration across the boundary may also play a role. Furthermore, this finding is in good qualitative agreement with studies that show the $(100)_m$ habit is the more favorable habit planes for related zirconia phase transformations \cite{guan_energy_2015}. Strain and displacements of the Hf  sub-lattice needed to form certain rumpled boundaries (i.e.~any boundaries with the $(001)_m$ plane) do not appear dramatically different than that of the observed boundaries, and so cannot be ruled out entirely.

\begin{figure}[H]
	\includegraphics[width=6.5in]{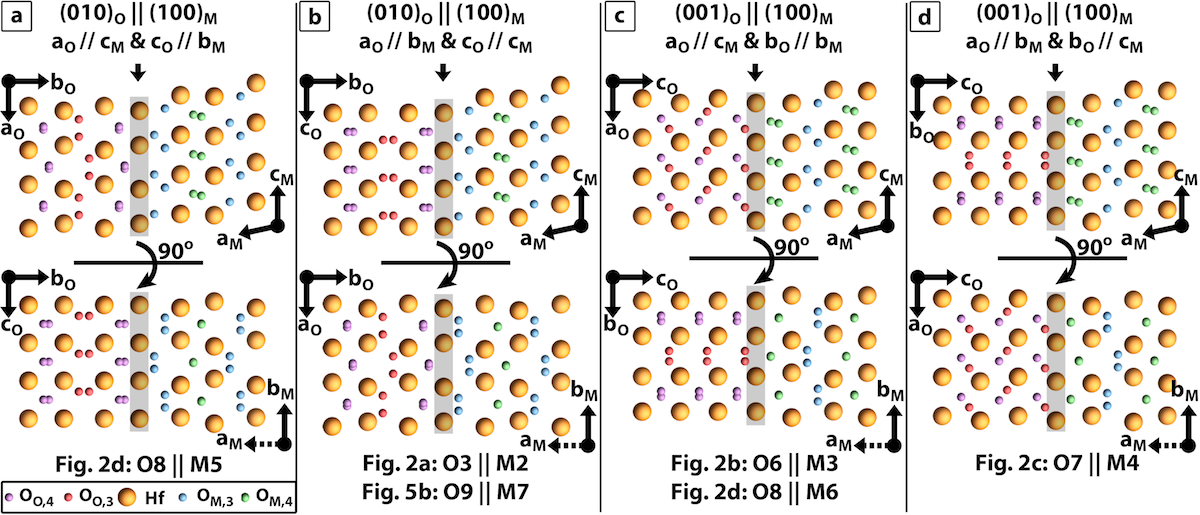}
	\caption{Schematics of the observed O/M (left/right) interphase boundaries. Top labels indicate the boundary  orientations. Bottom schematics are rotated 90$^\circ{}$ relative to the top view.  The image/phase pairs where these boundaries are found in the HAADF STEM images is listed beneath each schematic. The oxygen phase phase and coordination are also provided.}
	\label{fig:figure4}
\end{figure}



Approximate configurations of the structures observed at interphase boundaries are provided in Figure \ref{fig:figure4}. The shaded terminal Hf plane at the boundaries are those of the orthorhombic phase. Note that both phases have three-coordinate and four-coordinate oxygen atom positions. Because the oxygen sub-lattices are not observed in the STEM images, the schematics are approximate and represent one of several possible configurations. Nonetheless, the boundaries show how O/M boundaries can form that reasonably satisfy the symmetry of both phases.

The variety of observed interphase boundaries provides insight into how they might influence phase stabilization and enable phase transformation. Immobile boundaries artificially limit the grain size and impart a coherent strain onto the lattices, which is known to play an important role in phase stabilization \cite{materlik_origin_2015,batra_factors_2017,barabash_ferroelectric_2017}. Furthermore, these boundaries can influence domain pinning. A boundary capable of moving under the influence of an electric field would alter the electrical behavior by changing the phase fractions during cycling. For example, monoclinic/orthorhombic phase transformations have been initiated during electron beam irradiation in both zirconia particles \cite{chiao_martensitic_1990} and ceramics \cite{muddle_phase_1988}. While the boundaries observed here did not move during STEM imaging, identical boundary orientations exhibit several different configuration. 

\begin{figure}[H]
	\includegraphics[width=3.25in]{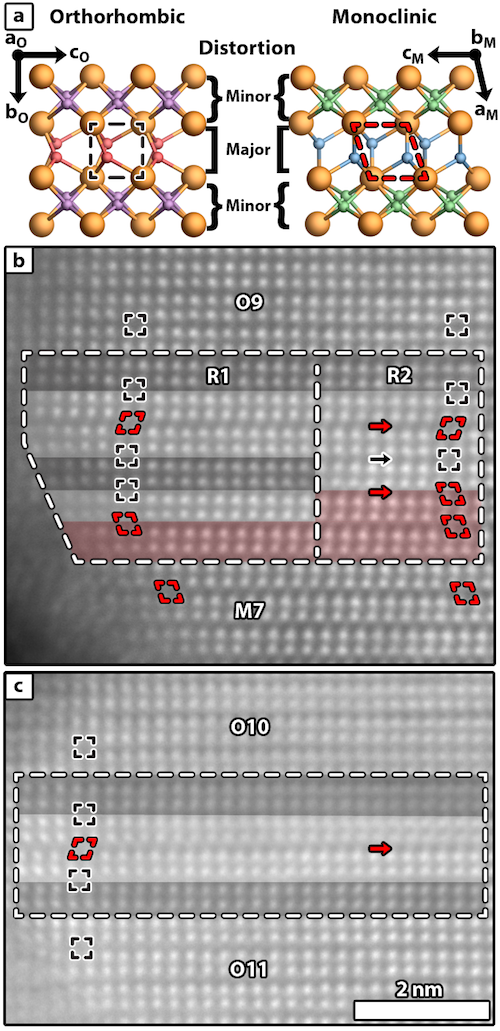}
	\caption{(a) Schematics of the O- and M-phases with the major and minor distortion half-units indicated. (b) A complex interphase boundary with color-coded symmetry overlays (black:orthorhombic, red:monoclinic, and white:both), and important major distortion units indicated. Arrows highlight key structural figures described in the main text, and apparent symmetry of the distortion units is mapped. (c) A similar boundary dividing two pure O-phase regions.}
	\label{fig:figure5}
\end{figure}


With respect to boundary mobility, Figure \ref{fig:figure5}b is instructive. This $(010)_o\|(100)_m$ boundary is oriented equivalently to ones previously shown to be mobile in particles of the related zirconia crystal structure \cite{chiao_martensitic_1990}. This boundary also displays many unique features that set it apart from the similar boundary presented earlier in Figure \ref{fig:figure2}a. Distinct orthorhombic O9 and monoclinic M7 regions are separated by a complicated interphase boundary, which is segmented into regions R1 and R2 by the dotted lines within the boundary region in Figure \ref{fig:figure5}a. The qualitative symmetry of the Hf sub-lattice is indicated by colored overlays with black indicating pure orthorhombic symmetry, red representing pure monoclinic symmetry, and white labeling regions where the Hf sub-lattice seemingly co-satisfies the symmetry of each phase. 

The monoclinic and orthorhombic unit-cells can be thought of as  distorted tetragonal unit-cells  \cite{trolliard_martensitic_2011}.  Half the the structure resembles, with minor distortion, the parent tetragonal phase, while the other half deviates significantly for both monoclinic and orthorhombic cells, see Figure \ref{fig:figure5}a.  The layers with minor distortions are structurally very similar between the monoclinic and orthorhombic phases. By contrast, the majority of the differences between the monoclinic and orthorhombic phase occur within the major distortion layers. 

The nominal structures of the major distortion layers are labeled in Figure \ref{fig:figure5}b. Several interesting structural features occur in this boundary. First, the bottom red arrow in R2 indicates a monoclinic-like major distortion layer that becomes an orthorhombic-like major distortion layer across the R2/R1 boundary. Next, the monoclinic major distortion layer at the same red arrow appears ``twinned'' with respect to the monoclinic major distortion layer indicated by the top red arrow. This twin-like feature occurs across a major distortion layer resembling the orthorhombic phase (black arrow). The monoclinic-like major distortion layer indicated by the top red arrow occurs between two major distortion units with orthorhombic-like structures. From this analysis, it becomes apparent that substitution of a monoclinic-like major distorted layer into an orthorhombic lattice results in an anti-phase-like boundary in the orthorhombic phase (see top red arrow in Figure \ref{fig:figure5}b and red arrow in Figure \ref{fig:figure5}c). Similarly, insertion of an orthorhomibic-like major distortion layer into the monoclinic lattice results in a ``twin-like'' defect (see black arrow in Figure \ref{fig:figure5}b).




The large variation in the structure at these boundaries hints at the potential for  mobility during the application of an electric field. Specifically, the ``snapshot'' in Figure \ref{fig:figure5}b suggests an interphase boundary in various states of converting between the monoclinic and orthorhombic lattices, much as suggested by the ``step-flow''-like motion of the same boundary orientation in Ref.~\cite{chiao_martensitic_1990}. Crystal chemistry suggests a double $\left[001\right]_o(010)_o$ glide system that converts the orthorhombic phase into the monoclinic phase, with the reverse occurring by a $\left[001\right]_m(100)_m$ double glide \cite{trolliard_martensitic_2011}. Such glides impact the symmetry of the entire unit-cell, though a majority of the structural changes occur within the more major distortion portions of the unit cells \cite{trolliard_martensitic_2011}. Consistent with this, the structure of the major distortion layer indicated by the bottom red arrow transitions between glide states at the boundary between regions R2 and R1. Furthermore, this glide system explains how insertion of one phase's major distortion structure into the other phase initiates features akin to ``anti-phase-like'' and ``twin-like'' defects, as discussed above.

Consistent with the understanding that some interphase boundaries are likely immobile in ferroelectric hafnia, M/O interphase boundaries are still observed after field-cycling, of which Figure \ref{fig:figure5}b-c is such an example. Internal discontinuities and strains as encountered near interphase boundaries are important for phase stabilization as they limit grain size and exert an internal force. Domain boundaries have been suggested as stabilizing higher symmetry phases in zirconia nanoparticles \cite{liu_metastable_2014}, and interphase boundaries can play a similar role in this system. Phase stability can change in the vicinity of interphase boundaries due to differences in local epitaxial strain \cite{materlik_origin_2015,batra_factors_2017,barabash_ferroelectric_2017}, or even due to a departure from the undistorted monoclinic and orthorhombic lattices \cite{clima_identification_2014,barabash_ferroelectric_2017}. In these instances, application of an electric field may be insufficient to destabilize one phase with respect to one another. Such boundaries would also play a role in fatigue mechanisms in these materials. 
 
Unlike the M/O interphase boundaries, no clear examples of 90$^\circ{}$ domains in the orthorhombic phase are found in the current work for field-cycled samples. Within the limits of the STEM sampling, this suggests that field-cycling results in increased domain uniformity by aligning some of the ``as-grown'' 90$^\circ{}$ domains. Such an  increase in domain uniformity would concomitantly increase the remanent polarization, which is  observed during wake-up when field cycling \cite{schenk_complex_2015,pesic_physical_2016,grimley_structural_2016}. Moreover, martensitic phase changes between high symmetry phases and the non-orthogonal monoclinic phase necessitates a shape change. This has been seen in the case of both orthorhombic zirconia particles \cite{chiao_martensitic_1990} and tetragonal hafnia-zirconia nanoparticles \cite{tang_martensitic_2005}. Twin formation in the monoclinic phase has been shown to minimize shear strain during such a transformation for both hafnia-zirconia nanoparticles \cite{tang_martensitic_2005} and hafnia thin films \cite{maclaren_texture_2009}. Due to a restricted geometry, thin films have fewer degrees of freedom by which to change shape, and likely rely more on generation of accommodating defects like dislocations and twin and/or anti-phase boundaries to convert between phases. Moreover, the shear strains required for such a transformation may be inaccessible to certain regions of the sample, locking in a higher symmetry phase \cite{maclaren_texture_2009}. As such, the geometric constraints due to electrode(s), neighboring grain(s), and/or other boundaries may immobilize some of the interphase boundaries with little to no room to move around their eccentric positions.

\section{Conclusions}

This work demonstrates the rich structural chemistry accessible to  ferroelectric HfO$_2$, which enables formation of a complex mixture of domains, planar defects, and interphase boundaries. The complex structure near interphase boundaries hints at a possible continuum between orthorhombic and monoclinic phases in the vicinity of the boundary walls. Further, the distortions present near these boundaries suggests the potential for  mobility. These insights yield new perspectives for the modeling of switching and domain wall motion, and provide a basis for comparison to domain wall and interphase boundaries in conventional perovskite ferroelectrics. Overall, this work lays the groundwork for calculations aiming to explore interphase boundary energetics, where further knowledge is needed to improve stability, mobility, and their impact of field-cycling.


\section{Methods}

\subsection{Sample Information}

\noindent 27 nm Gd:HfO$_2$ capacitors with 10 nm TiN electrodes were grown using atomic layer deposition as described previously \cite{hoffmann_stabilizing_2015}.  Lamella were prepared for scanning transmission electron microscopy (STEM) by focused ion beam from both cycled and pristine devices using an FEI Quanta. Details of the field cycling can be found in  Ref.~\cite{grimley_structural_2016}.



\subsection{Scanning Transmission Electron Microscopy}

\noindent High-angle annular dark-field (HAADF) STEM was performed on an FEI Titan G2 60-300 kV equipped with a probe-corrector and an X-FEG source. The microscope was operated at 200 kV with a detector inner semi-angle of approximately 77 mrad, probe currents of around 80 pA measured with the current monitor on the screen, and probe semi-convergence angle approximately equal to 19.6 mrad. RevSTEM images \cite{sang_revolving_2014} were  acquired using 40 1024 x 1024 pixel frames with a 2 $\mu{}$s/pixel dwell time and a 90$^{\circ}$ rotation between each successive frame. Where necessary, scan coil distortion was removed by previously described methods  \cite{dycus_accurate_2015}. The atom column positions were determined by fitting  two-dimensional Gaussian distributions via MATLAB scripting \cite{sang_atom_2014}. PACBED patterns were simulated using the MBFIT (``Many-Beam dynamical-simulations and least-squares FITting'') package by K. Tsuda at Tohoku University \cite{tsuda_refinement_1999}. Simulation parameters matched those from experiment.  Structural parameters were taken from Refs.~\cite{kisi_crystal_1989, ohtaka_synthesis_1991, ohtaka_structural_1995,  sang_structural_2015}. The simulation output was rescaled using bicubic interpolation to match experiment.


\acknowledgement
Christoph Adelmann from Imec, Belgium is gratefully acknowledged for depositing the TiN-Gd:HfO2-TiN stacks. The authors thank Jacob L.~Jones for helpful feedback and discussions. EDG and JML gratefully acknowledge support from the National Science Foundation (Award No. DMR-1350273). EDG acknowledges support for this work through a National Science Foundation Graduate Research Fellowship (Grant DGE-1252376). TS, US~and TM~gratefully acknowledge the German Research Foundation
(Deutsche Forschungsgemeinschaft) for funding part of this
research in the frame of the ``Inferox'' project (MI 1247/11-2). This work was performed in part at the Analytical Instrumentation Facility (AIF) at North Carolina State University, which is supported by the State of North Carolina and the National Science Foundation (award number ECCS-1542015). The AIF is a member of the North Carolina Research Triangle Nanotechnology Network (RTNN), a site in the National Nanotechnology Coordinated Infrastructure (NNCI). 
\suppinfo

\bibliography{Interphase_Boundaries.bib}

\end{document}